\documentclass[aps,prl,twocolumn,secnumarabic,amsppt,superscriptaddress,showpacs]{revtex4}
\usepackage{graphicx}
\pdfoutput=1
\usepackage{amsmath,amsfonts,amssymb}
\usepackage{wrapfig}
\usepackage{bbm}
\usepackage[usenames]{color}
\makeatletter

\begin{document}
\title{An atomic interface between microwave and optical photons} 

\author{M. Hafezi }
\altaffiliation[email: ]{hafezi@umd.edu}
\affiliation{Joint Quantum Institute, University of Maryland/National Institute of Standards and Technology, College Park, MD 20742}
\author{Z. Kim}
\affiliation{Joint Quantum Institute, University of Maryland/National Institute of Standards and Technology, College Park, MD 20742}
\author{S. L. Rolston}
\affiliation{Joint Quantum Institute, University of Maryland/National Institute of Standards and Technology, College Park, MD 20742}
\author{L. A. Orozco}
\affiliation{Joint Quantum Institute, University of Maryland/National Institute of Standards and Technology, College Park, MD 20742}
\author{B. L. Lev}
\affiliation{Departments of Applied Physics and Physics, and E. L. Ginzton Laboratory, Stanford University, Stanford, CA 94305}
\author{J. M. Taylor}
\affiliation{Joint Quantum Institute, University of Maryland/National Institute of Standards and Technology, College Park, MD 20742}

\begin{abstract}
A complete physical approach to quantum information requires a robust interface among flying qubits, long-lifetime memory and computational qubits.  Here we present a unified interface for microwave and optical photons, potentially connecting engineerable quantum devices such as superconducting qubits at long distances through optical photons.  Our approach uses an ultracold ensemble of atoms for two purposes: quantum memory and to transduce excitations between the two frequency domains.  Using coherent control techniques, we examine an approach for converting and storing quantum information between microwave photons in superconducting resonators, ensembles of ultracold atoms, and optical photons as well as a method for transferring information between two resonators. \end{abstract}

\maketitle

Controlling the interaction between quantum bits and electromagnetic
fields is a fundamental challenge underlying quantum information
science. Ideally, control allows storage, communication, and
manipulation of the information at the level of single
quanta. Unfortunately, no single degree of freedom satisfies all these
criteria simultaneously \cite{Ladd:2010p27648}.  Instead, a hybrid approach may
take advantage of each system's most attractive
properties. For example, optical photons provide a robust
long-distance quantum bus~\cite{Obrien:2009p27731}, while microwave
(MW) photons can be easily manipulated using superconducting qubits
\cite{Schoelkopf:2008p8712}, and atoms can store quantum information
for seconds or even minutes~\cite{lukin2,fleischhauer1}.  We
propose an interface between optical, microwave photons and atomic
excitations that takes advantage of each of these properties.

\textcolor{black}{Previous proposals for interfaces of this nature considered magnetic coupling
between ultracold atoms
\cite{Verdu:2009} or spins \cite{Imamoglu:2009,Marcos:2010, Wu:2010, Kubo:2010,Amsuss:2011}  to superconducting waveguides, or using opto-mechanics for frequency conversion between optical and MW photons, without providing a medium for storage \cite{Tsang:2010,Regal:2011,Taylor:2011}.}
In a recent proposal, the possibility of coupling
ultracold atoms to a nanofibre in the vicinity of a superconducting
waveguide/resonator was suggested [Fig.1(a)] \cite{Hoffman:2011}.
The evanescent tail of the two-color laser field propagating in the
fiber provides the necessary potential to trap atoms close to the
nanofibre in the form of a 1D lattice, which has recently been demonstrated in Ref.~\cite{Vetsch:2009}.  Exponential decay of the optical trapping field allows the atoms to be held close to the superconducting waveguide, leading to a large atom-photon
coupling in the microwave domain.  Furthermore, the nanofibre provides
an optical waveguide both for trap light and for optical access to
atoms.  Such a system would provide a simultaneous interface between optical
and MW photons and atomic ensemble quantum memory.
\begin{figure}[t]
\includegraphics[width=0.44\textwidth]{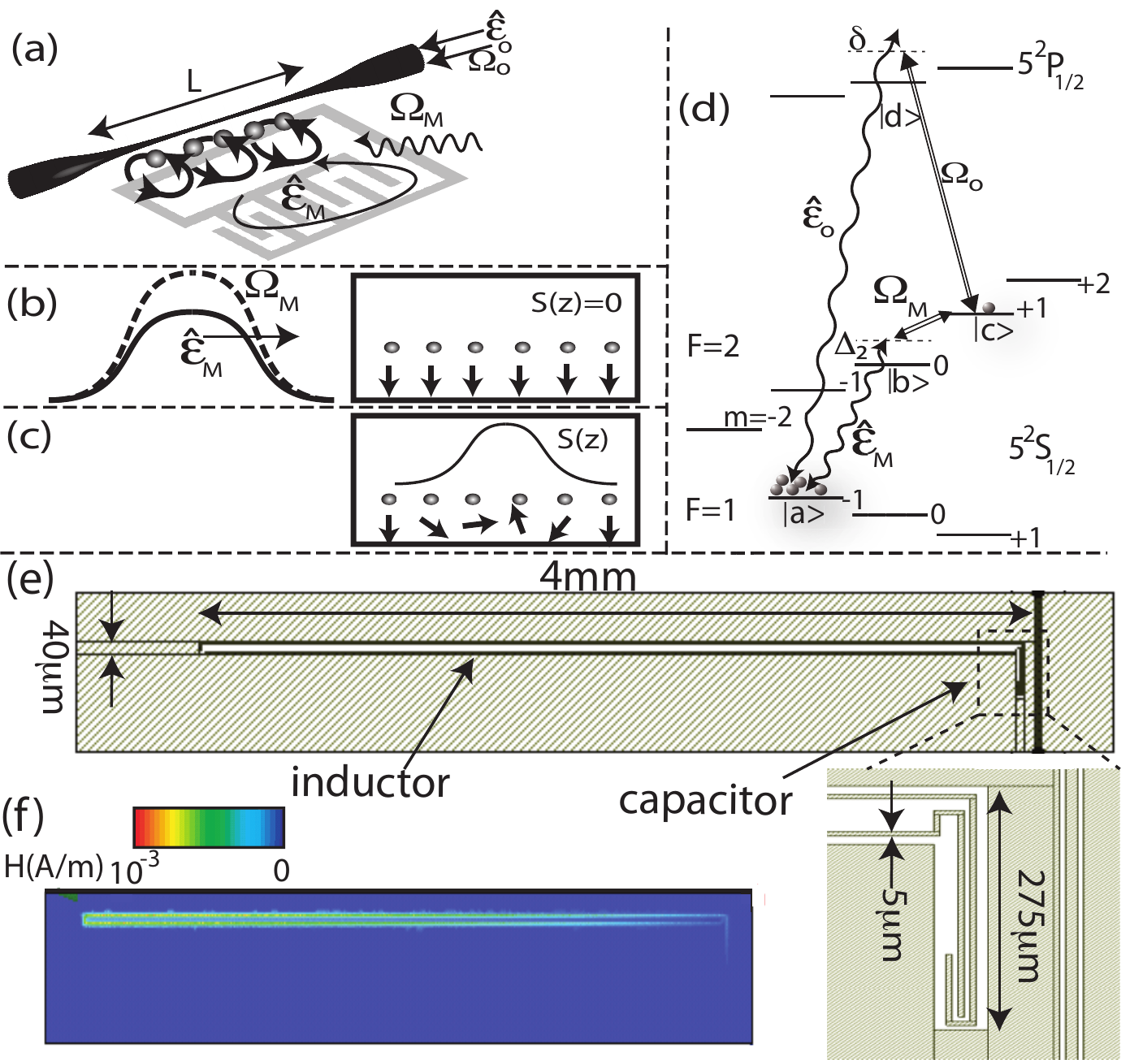}
\caption{(a)
  Schematics of the interface: Atoms (sphere), in a 1D lattice of the length L, are electrically (magnetically)
  coupled to light in the nanofiber (the superconducting waveguide),
  respectively. The quantum microwave (optical) field with Rabi frequency $\hat{\mathcal{E}}_{\rm{M}}$ ($\hat{\mathcal{E}}_o$) is manipulated using a classical control radio-frequency (optical) field with Rabi frequency $\Omega_{\rm{M}}$ ($\Omega_o$), respectively.  Trapping lights are not shown in the figure. (b) The quantum ($\hat{\mathcal{E}}_{\rm{M}}$) and control ($\Omega_{\rm{M}}$) electromagnetic fields
  arrive while the atomic system is in the ground state. (c) The quantum
  field is stored as an atomic spin excitation ($S(z)$). (d) Internal level
  structure of a $^{87}$Rb atom and transitions induced by the four electromagnetic
  fields. $\delta,\Delta_2$ are two-photon and one-photon detunings, respectively. (e) Dimensions of an example LC resonator. (f) Magnetic field profile viewed from the top, lighter colors show higher fields.}
\end{figure}

In this Letter, we theoretically illustrate how this proposed system enables
storage, retrieval and conversion for optical and MW photons.
Experiments have established
\cite{Julsgaard:2004,Chaneliere:2005,Eisaman:2005,Chou:2005} that
coherent control techniques of multilevel atoms
\cite{lukin2,fleischhauer1}, based on electric-dipole coupling in a
$\Lambda$-system, allow for efficient storage and retrieval of optical
photons into atomic excitations from an ensemble of atoms.  We adapt
this approach to also store and retrieve MW photons.  We investigate
the effect of finite bandwidth of MW photons on the
storage-retrieval process and also the effect of periodicity of the
atomic ensemble which can change the propagation of optical photons
due to Bragg scattering.  We conclude by discussing quantum
communication and measurement protocols enabled by our interface as well as the use of non-alkali atoms for enhanced coupling~\cite{lu:2010}.

The storage of photons (either in the MW or optical domain) in
atomic excitations in a generic $\Lambda$-system forms the basis for
our interface, and is shown in Fig.~1.  Specifically, we start with an
optically pumped ensemble of $N$ atoms in one of the hyperfine ground
states, $|a\rangle$. A classical control field $\Omega_{\rm{M}}(t)$ (or
$\Omega_{o}(t))$ is used to coherently manipulate the coupling between
an intermediate state $|b\rangle$ (or $|d\rangle$) and a final ground
state $|c\rangle$.  These control fields in turn determine the
propagation of the quantum field $\hat{\mathcal{E}}_{\rm{M}}(t)$ (or
$\hat{\mathcal{E}}_{o}(t)$) coupling $|a\rangle$ to $|b\rangle$
($|d\rangle$), leading to  electromagnetically induced
transparency (EIT) and slow light. The evolution of such coupled
system is best described by a bosonic dark state polariton
\cite{Fleischhauer:2000kx}, with creation operator:
\begin{equation}
  \hat{\Psi}_i^{\dagger}(z,t)=\frac{\Omega_i(t)\hat{\mathcal{E}}_i^{\dagger}(z,t)-g_i\sqrt{N}\hat{S}^{\dagger}(z,t)}{\sqrt{\Omega_i^{2}(t)+g_i^{2}N}}\label{eq:polariton}\end{equation}
where $i$ corresponds to either optical (O) or  MW domains (M). $\hat{S}^{\dagger}(z,t)$ is spin wave creation operator associated with the atomic ground state coherence $|c\rangle\langle a|$, in the continuum limit \cite{Fleischhauer:2000kx,gorshkov:storage} --- we discuss the effect of lattice later in the Letter.  Here  $g_o(g_M)$ is the electric  (magnetic) dipole coupling between the atoms
and the optical (microwave) waveguide photons, respectively, and $N$ is
the total number of atoms. 

During the entire operation,  quantum excitations remain in the form of a dark polariton and the ratio between the enhanced atom-photon coupling and the control field ($ \eta_{i}=g_{i}\sqrt{N}/\Omega_{i}$) dictates the
mixture between atomic and photonic parts. In particular, when the
control is strong $( \eta_i \ll1)$, the polariton is
mostly photonic and the system is transparent, with a group velocity
near to the speed of light. In contrast, when the control is weak
$(\eta_i \gg1$), the polariton is mostly atomic, with a group velocity
approaching zero (slow light). Changing the control field
allows one to transform a photonic excitation into an atomic
excitation and vice versa (see Ref. \cite{lukin2,fleischhauer1}): once
the incoming pulse is entirely inside the system, the control field is
adiabatically turned off ($\Omega_i(t)\rightarrow0$), and the excitation
will be stored as atomic spin excitations in the ground state manifold
($S^{\dagger}(z,t)$) [Fig.~1(b,c)].  By reversing the control field(s)
in time, the stored atomic excitations can later be retrieved as MW or
optical photons.


We first address the practical challenges for storing and retrieving
MW photons. As an example case, we consider trapped $^{87}$Rb atoms
coupled to the MW waveguide through the magnetic-dipole interaction,
which is characterized by single-photon Rabi frequency $g_M$ \cite{Verdu:2009,Hoffman:2011}. The optimal states to preserve the ground state coherence are the
clock states \cite{Treutlein:2004,Zhao:2008}:
$|a\rangle=|F=1,m_{f}=-1\rangle$ and $|c\rangle=|F=2,m_{f}=+1\rangle$,
as shown in Fig.~1(d).  The ground state decoherence rate is dominated
by off-resonant scattering of photons from the trapping lasers which
is relatively small $\gamma\simeq 20~\rm{s}^{-1}$ \cite{Vetsch:2009}. The
quantum field ($\hat{\mathcal{E}}_{\rm{M}}$) and the classical
radio-frequency control field ($\Omega_{\rm{M}}$) are shown in
Fig.~1(d). In order to isolate a single atomic transition to interact
with the MW photon, we can use polarization and frequency
selectivity. Furthermore, the application of a moderate magnetic field ($\simeq 6$ mT),  leads to a necessary quadratic Zeeman shift and makes the multilevel corrections negligible.  For example, as shown in Fig.~1(d), a Zeeman field can
split the degeneracy within the hyperfine states so that only the two-photon
transition  between $|a\rangle$ and $|c\rangle$ is possible.

We review the conditions under which the standard optical EIT storage
technique is efficient \cite{lukin2,fleischhauer1,gorshkov:storage} and
apply them to the microwave domain. We focus on the free-space case,
and the generalization to the resonator case can be done by replacement:
$c/L\rightarrow\kappa$, the resonator decay rate. The bandwidth of the incoming (or retrieval) photon cannot be
arbitrarily large. Analytical  and numerical calculations have shown that an adiabatic
condition must be fulfilled \cite{gorshkov:storage}, and photons with large bandwidth can
not be stored since the system does not have fast enough response
time. This condition can be intuitively derived, as discussed in
Refs.~\cite{gorshkov:storage}. Briefly, a single spin-wave excitation transfers into the waveguide
with a rate $\frac{g_M^{2}N}{c/L}$ and decays via a decoherence rate
$\gamma$. Consequently, via
time-reversal symmetry, the bandwidth of an incoming photon to be
stored must satisfy $T_{p}^{-1}\leq\frac{g_M^{2}N}{c/L}$, where $T_p$ is the pulse duration. The retrieval efficiency is equal to the ratio between the transfer rate and the combined transfer and decay rates, i.e. $1-\frac{\gamma c/L}{g_M^2 N}$. In the optical domain, this efficiency is simply given by the optical depth of the system, i.e. $\simeq 1-\frac{\Gamma_{tot} c/L}{4\pi g_o^2 N}$ \cite{gorshkov:storage}, where $\Gamma_{tot}$ is total spontaneous emission rate of the optical transition and where $g_o$ is the single-photon electric-dipole coupling to the fiber.

For a given pulse duration, which satisfies the maximum bandwidth condition (above), the medium should initially be transparent to the pulse
$(T_{p}^{-1}\ll\Delta\omega_{EIT})$, where $\Delta\omega_{EIT} $ is the width of the transparency
window.  At the same time, the entire pulse should fit inside the medium:
$T_{p}v_{g}\preceq L$. Given that the group velocity $v_{g}/c=1/(1+\eta_{M}^2)\simeq
\Omega_{M}^{2}/g_M^{2}N$ and $\Delta\omega_{EIT}\simeq\frac{\Omega^{2}_{M}}{\gamma}\sqrt{\frac{\gamma
    c/L}{g_M^{2}N}}$, the last two conditions can only be satisfied in the high cooperativity limit:
$\frac{g_M^{2}N}{\gamma c/L}\gg1$. In the optical EIT schemes, where
the transition decay is dominated by radiative decay, the condition is
equivalent to requiring large optical depth. Moreover, the required control field is optimal when the bandwidth of the incoming photon matches the reduced resonator linewidth (due to a slow light effect) \cite{gorshkov:storage},  i.e., $T_{p}^{-1}\simeq\frac{\Omega^{2}_{M}\kappa}{g_M^{2}N}$.

The bandwidth of MW photons originating in the resonator is
given by the resonator bandwidth ($T_{p}^{-1}\simeq\kappa$). We consider a LC resonator, schematically shown in the Fig.~1(a), where the inductor part is long enough ($\simeq 4$ mm) to accommodate many atoms  ($N$ = 8,000).  Using simulation software, we tune the capacitor to achieve a resonance with $^{87}$Rb around  $\omega_{MW}/2\pi=6.8$ GHz  (for details see Ref.~\cite{Hoffman:2011}). The magnetic field is relatively uniform along the inductor part, as shown in Fig.~1(f). The corresponding single-photon magnetic coupling is estimated numerically to be  $g_M/2\pi= 70$ Hz. Assuming a quality factor $Q\simeq 10^{6}$, $\kappa\simeq 43~\rm{ms}^{-1}$, the bandwidth condition can be satisfied. Under these conditions, the magnetic cooperativity  is $\frac{g_M^{2}N}{\gamma \kappa}\simeq 1700$. The required RF control field should be $\Omega_{\rm{M}}/2\pi\simeq 6.3$ kHz. The optical coupling is
characterized by the ratio between the spontaneous emission into the
fiber and the total spontaneous emission, i.e.,
$\Gamma_{\rm{wg}}/\Gamma_{\rm{tot}}=4\pi
  g_o^2 L/\Gamma_{\rm{tot}} c$. This ratio is estimated to be
$5\%$ \cite{Hoffman:2011}.  Therefore, the electrical cooperativity (optical depth) of the system is OD$=\!N\Gamma_{\rm{wg}}/\Gamma_{\rm{tot}}\!\simeq 400$. The high electric (magnetic) cooperativity guarantees efficient transfer of excitations in the optical (MW) domain, respectively.

We now  shift our attention to the new features of the optical components of our
interface.  Trapped $^{87}$Rb atoms are coupled to light in the optical fiber
through the electric-dipole interaction. Since the atoms are trapped in a one-dimensional lattice along the
optical fiber, the light propagating inside the fiber experiences
periodic scattering in the form of a Bragg grating. The effect of such
multiple scatterings can lead to a bandgap structure, in direct
analogy to two-level atoms in an optical lattice
\cite{Deutsch:1995}. Since the atoms are not saturated  by small numbers of 
 photons $(\Gamma_{\rm{wg}}/\Gamma_{\rm{tot}}\ll1)$, the system
 is linear, i.e., away from the photon-blockade regime \cite{Chang:2007,ShanhuiFan:PRL2007}. As the atoms are periodically spaced, we can
discretize the propagation of the electric field (Fig.~2(a)) and use the
transfer matrix formalism to study these multiple scatterings, due to three-level transition ($|a\rangle \leftrightarrow |d\rangle \leftrightarrow |c\rangle$)(see the Supplemental Material for details).

The transmission spectrum of a 1D array of 8000 sites ($=L/a$) is shown in
Fig.~\ref{fig:band_theory}(b). We find that our realistic numbers lead to
behavior close to that of free space and no band gap structure is observed
\cite{Petrosyan:2007,Nunn:2010,Witthaut:2010}. In particular, the dips in the
transmission spectrum correspond to dressed states split by the Rabi
frequency of the control field, in direct analogy to EIT in free
space. In the middle of the
transparency window, the excitations transform entirely into spin
excitation of the atoms. Therefore, once the pulse is inside the
atomic medium, by turning off the control field, the photonic
excitation can be stored as atomic ground state excitation.  
Deviations from this behavior can occur, particularly for stronger single atom-field coupling, 
but this 
will be the subject of future research.

\begin{figure}[t]
\includegraphics[width=0.45\textwidth] {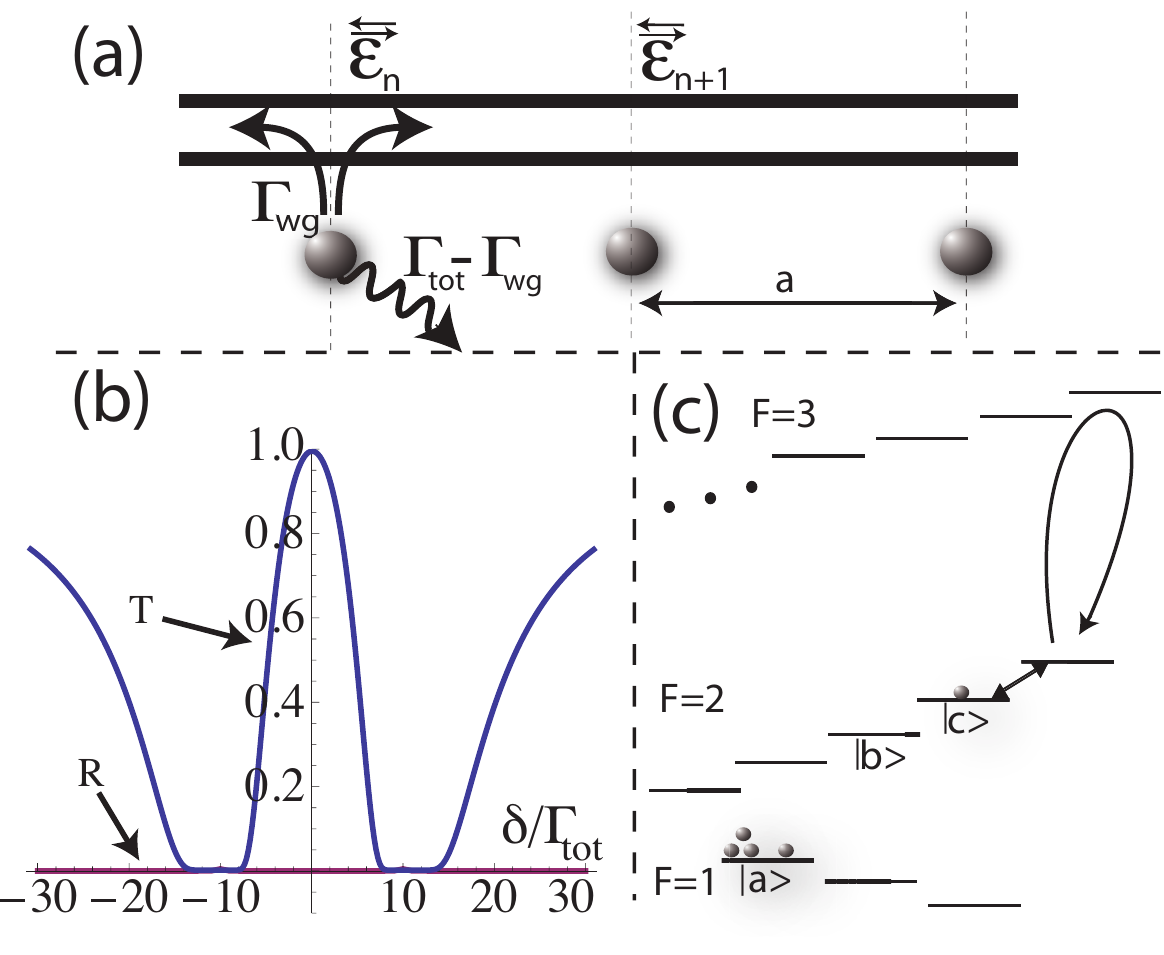}
\caption{(a) Atoms are periodically coupled to the forward  and backward-going light, in the fiber. (b) The transmission (T) and reflection (R) spectrum of light due to interaction with 1D lattice of atoms. In this plots, $(\Gamma_{wg},\Omega,c/a)/\Gamma_{tot}=(0.05,10,10^7)$,  the lattice spacing $a=500$ nm, and one atom per site is considered. (c) Atomic level configuration for  detecting a single excitation.} 
\label{fig:band_theory}
\end{figure}
%

In such hybrid system, we can implement various protocols. First, one can
coherently transfer the optical photon, MW photons and atomic
excitation to each other. As mentioned earlier, this process is
efficient when the cooperativity is large $\frac{g_M^{2}N}{\gamma
  c/L}\gg1$ and the photon pulse duration satisfies
 $T_{p}^{-1}\leq\frac{g_M^{2}N}{c/L}$.   This enables a quantum-coherent interconnection between MW excitations
and optical photons, which allows for a wide variety of quantum
communication and quantum information protocols between distant
systems, including quantum repeaters, teleportation-based gates, and
distributed quantum computing \cite{Zeilinger:Book}.

Second, in this interface, single photons  can be detected with high quantum efficiency. In
particular, when the excitation is a photon (either
optical or MW), it can transferred to an atomic spin wave.  In turn, the atomic
spin wave can be transferred to a hyperfine excitation detectable by absorption \cite{Brask:2010}, as shown in
Fig.~2(c). More specifically, the single atomic excitation in form of
$|c\rangle\langle a|$ coherence can be efficiently transferred to
$|F=2,m_{F}=2\rangle\langle F=1,m_{f}=-1|$, using MW and RF control
fields \cite{Chaudhury:2007,Merkel:2008}.  Then, using the cycling
transition, we can verify the number of original excitations with a
high degree of confidence using, e.g., Bayesian inference.

Third, we can generate entanglement between a MW photon and the
atomic ensemble in analogy to off-resonant Raman atom-photon
entanglement generation \cite{duan1}.  The atomic ensemble should be
prepared at $|c\rangle=|F=2,m_{F}=+1\rangle$ level; then applying a RF field
coherently generates an anti-Stokes MW photon accompanied with a atomic
excitation in the coherence $|F=2,m_{F}=+1\rangle\langle F=1,m_{F}=-1|$, where the outgoing MW photon and the atomic ensemble
are entangled. The efficiency of this process is similar to the
storage-retrieval of single excitations, as discussed earlier. 

Fourth, we can envisage using the hybrid system to induce a large
optical nonlinearity via known Josephson junction-based microwave
nonlinearities. In particular, when the control fields
$\Omega_{o},\Omega_{\rm{M}}$ are on, through a four-wave-mixing process,
the optical and MW photons will be coupled to each other.  Therefore,
by adding a nonlinear element for MW photons (e.g., Cooper pair boxes
or superconducting qubits \cite{Vion:2002,Chiorescu:2003p33897}), a
large optical nonlinearity can be induced for optical photons. Such a large nonlinearity could be harnessed to perform a two-qubit phase
gate on optical photons, a key ingredient of deterministic 
optical quantum computing.

\begin{figure}[t]
\includegraphics[width=0.45\textwidth]{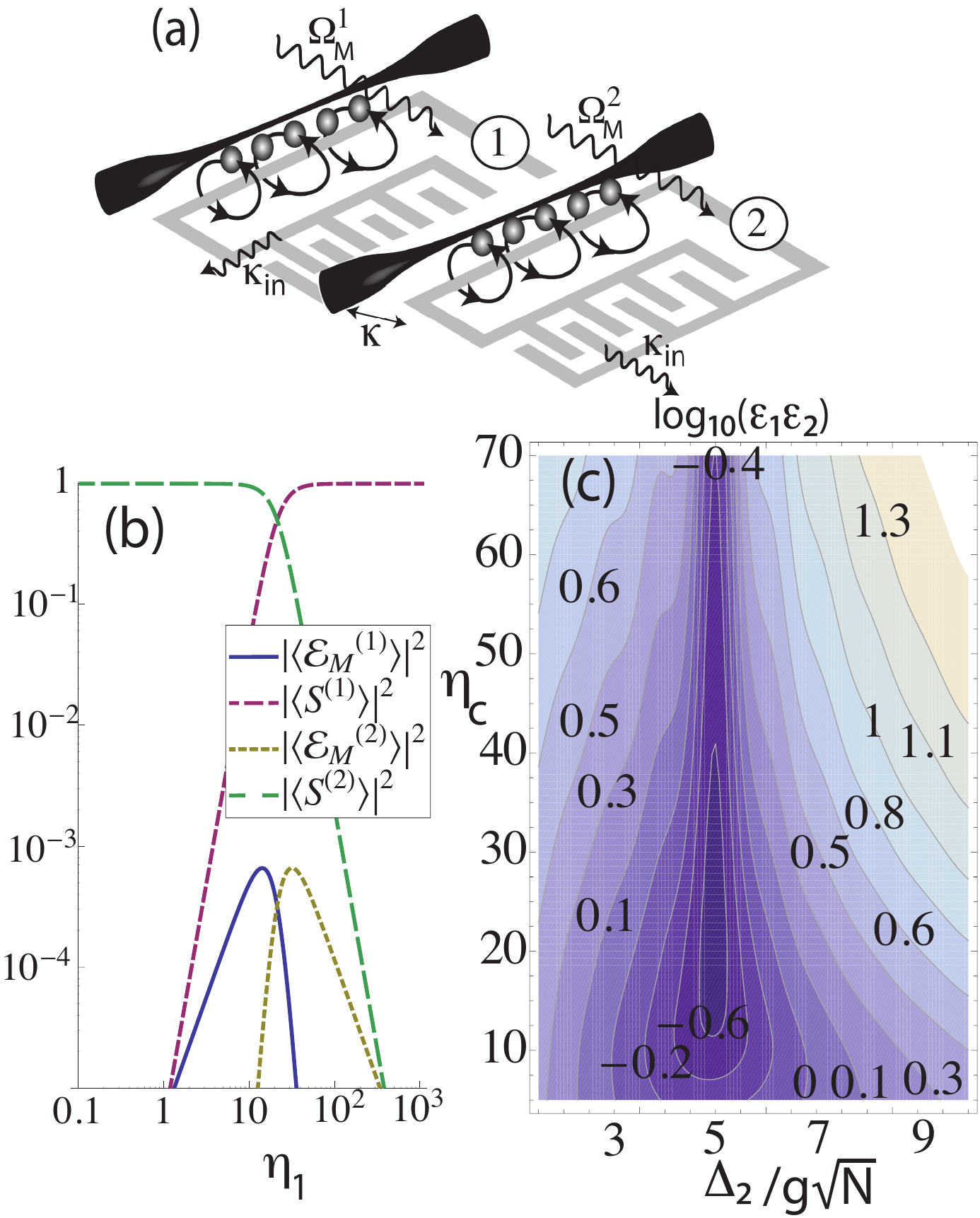}
\caption{ (a) Schematics of  two cells (two LC resonators  are coupled by the rate $\kappa$ and each having intrinsic loss $\kappa_{in}$), an adiabatic quantum state transfer can be performed, using atomic ensembles. (b) Probabilities of having the excitation in the photonic (normalized $|\langle\mathcal{E}^{(1)}_M\rangle|^2,|\langle\mathcal{E}^{(2)}_M\rangle|^2$) or atomic (normalized $|\langle S^{(1)}\rangle|^2,|\langle S^{(2)}\rangle|^2$) form, as a function of control field $\eta_1$, where $\eta_2=\eta_c^2 \eta_1^{-1}$, for optimized values of $(\eta_c,\Delta_2)\simeq(20,5)g\sqrt{N}$. (c) The combined adiabaticity-loss condition for crossing value of control fields ($\eta_c$) and one-photon detuning ($\Delta_2$), as shown in Fig.~1.  }
\label{state_transfer}
\end{figure}

Finally, we can use this system to coherently transfer quantum
excitations between two coupled cells, each comprising a resonator and atomic ensemble
(Fig.~\ref{state_transfer}(a)) \cite{Cirac:1997, VanEnk:1999,
  Pellizzari:1997, Lukin:2000, Vitanov:2001}, where the coupling rate is $\kappa$ and the intrinsic resonator loss rate is $\kappa_{in}$. By changing the control field ($\Omega_{M}$) in
each cell, one can dynamically control the resonator decay rate,
i.e. $\kappa/\sqrt{1+\eta^2}$ and perform dynamical impedance matching \cite{Lukin:2000}. If $\eta_1\ll 1\ll
\eta_2$, then the dark state in cell one (two), is mostly photonic
(atomic), respectively. Therefore, by adiabatically going from
$\eta_1\ll \eta_2$ to $\eta_2 \ll \eta_1$, we can transfer an atomic
excitation from cell two to cell one. The process can be performed
with high fidelity because the photonic mode is never excited and the
system remains in a dark state, as shown in Fig.\ref{state_transfer}(b),
where $\eta_2=\eta_c^2 \eta_1^{-1}$ and the control field values cross
at $\eta_c$ (see the Supplemental Material for details). During this process, two conditions should be satisfied:
(1) Adiabaticity: $\sum_{i\neq4} \frac{|\langle
  e_4|\partial_\eta|e_i\rangle|}{e_4-e_i}\dot\eta =\epsilon_1 \dot\eta
\ll 1$ where $|e_i\rangle$ are energy eigenstates of the system and
$e_i$ their corresponding energies, the dark state of interest is
$|e_4\rangle$. (2) Negligible loss:  $\sum_{i=1,2}\int (\kappa_{in}
|\langle\mathcal{E}^{(i)}_M\rangle|^2+\gamma |\langle S^{(i)} \rangle|^2)
\frac{1}{\dot \eta}d\eta = \epsilon_2/ \dot \eta \ll 1$, where the
first (second) term represents the photonic (atomic) loss,
respectively. In
order to satisfy both, one should have $\epsilon_1\epsilon_2\ll1$ (see
Fig.\ref{state_transfer}(c)). For relevant experimental parameters
$(\kappa,\kappa_{in},\gamma)\simeq (0.2 ,1,.0005) g\sqrt{N} $, we find
the optimized values for the crossing value of control fields and
one-photon detuning to be $(\eta_c,\Delta_2)\simeq(20,5)g\sqrt{N}$,
which makes $\epsilon_1\epsilon_2\simeq0.26$.

\textcolor{black}{We note that implementation of such schemes is within the reach of current technology, although it is challenging. In particular, the long nanofiber (a few cm) should be mounted in the proximity of the superconducting resonator without sagging or breaking \cite{Barclay:2006}. Moreover, the nanofiber polarization and transmission properties should be maintained during the transport into the dilution fridge and the cooling process to milli-Kelvin temperatures. The stray light scattered from the nanofiber can decrease the quality factor of the resonator, and therefore, using a material with higher $T_c$ such as TiN is preferred to Al \cite{Vissers:2010}.} While the coupling techniques and protocols proposed here have been presented for alkali atoms, they can be also implemented with rare earth elements.   In particular, for ultracold fermionic Dy~\cite{lu:2010}, one can benefit from 10$\times$-enhanced magnetic cooperativity in the RF regime, which, e.g., reduces practical constraints on atom number, while allowing coherent state transfer to both the telecom regime (1322~nm) and quantum dot transitions (953~nm, 1001~nm).  In summary, we have illustrated that an atomic ensemble coupled to an
optical and a microwave waveguide can serve as a long-lifetime memory
as well as a photon converter between microwave and optical
electromagnetic fields.  

This research was supported by the U.S. Army Research Office MURI award W911NF0910406,  ARO Atomtronics MURI  and by NSF through the Physics Frontier Center at the Joint Quantum Institute. We thank A. Gorshkov for fruitful discussions and E. Tiesinga and S. Polyakov for critical reading of the manuscript.

\bibliographystyle{apsrev}

\newpage

\begin{center}
{\Large Supplementary Material}
\end{center}

\section*{S1: Atoms on a one-dimensional lattice}

\begin{figure}
\includegraphics{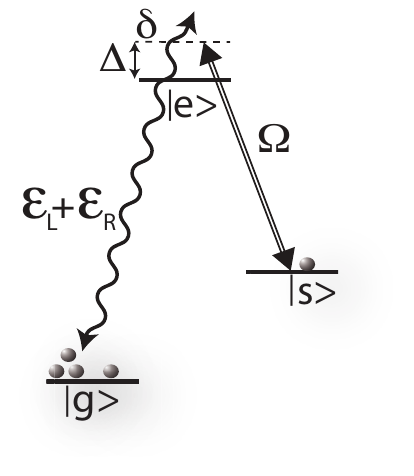}\caption{Three-level system coupled to an electromagnetic waveguide with left-
(right-) going field, $\mathcal{E}_L(\mathcal{E}_R)$, respectively. The control field
($\Omega)$ is detuned from e $\leftrightarrow$ s transition
by $\Delta$. \label{fig:Three-level-system-coupled}}
\end{figure}

\maketitle
Here we study the dispersive effect of atoms when they are trapped
in a one dimensional optical lattice, as expected for the tapered fiber trap given in the paper.  We consider a situation where  the atoms are not saturated with few photons  $(\frac{\Gamma_{\rm{wg}}}{\Gamma_{\rm{tot}}}\ll1)$, and therefore, the system is in the linear regime and the electric field operator can be presented by its expectation value, i.e., $\langle\hat{\mathcal{E}}(x)\rangle$. Furthermore, since the atoms are periodically spaced, we can discretize the propagation of the electric field (as shown in Fig.~2 of the main text) and use the transfer matrix formalism to study multiple scattering events to all order \cite{Deutsch:1995}. In the $n$th cell, we define the forward (backward)-propagating field as $\overrightarrow{\mathcal{E}}_{n}(\overleftarrow{\mathcal{E}}_{n})$, respectively. The fields in two consecutive cells are related by:
\begin{equation}
\left(\begin{array}{c}
\overrightarrow{\mathcal{E}}_{n+1}\\
\overleftarrow{\mathcal{E}}_{n+1}\end{array}\right)=M_{\rm{cell}} \left(\begin{array}{c}
\overrightarrow{\mathcal{E}}_{n}\\
\overleftarrow{\mathcal{E}}_{n}\end{array}\right)
\end{equation}
where $M_{\rm{cell}}$ is the transfer matrix and the corresponding transmission (reflection) coefficient is given by: $t=\frac{1}{M_{22}} ~(r=\frac{M_{12}}{M_{22}}) $, respectively. The transfer matrix of each cell is the product of two terms: $M_{\rm{cell}}=M_{\rm{free}}M_{\rm{atom}}$. The first term corresponds to the free propagation between two sites: 
\begin{equation}
M_{\rm{free}}=\left(\begin{array}{cc}
e^{ik_{p}a} & 0\\
0 & e^{-ik_{p}a}\end{array}\right),
\end{equation} 
where $k_{p}$ is the wave number of the incoming field, and $a$ is the lattice spacing. The second represents the light scattering due to the presence of atoms:
\begin{equation}
M_{\rm{atom}}=\left(\begin{array}{cc}
1+i\zeta & i\zeta \\
-i\zeta & 1-i\zeta\end{array}\right),\end{equation} 
where $\zeta$ characterizes the light scattering from the trapped atoms in a single site. In other words, the atomic transmission (reflection) coefficient is $t_a=\frac{1}{1-i\zeta} \left(r_a=\frac{i\zeta}{1-i\zeta}\right)$, respectively. Now, we study the scattering of photons from a single site (i.e. single atom) to find $\zeta$. A generic three-level system (Fig.\ref{fig:Three-level-system-coupled})
coupled to an electromagnetic waveguide can be described by the following
Hamiltonian \cite{Shen:05,Fan_PRA:2007,Chang:2007,Witthaut:2010}:

\begin{equation}
H=H_{atom}+H_{field}+H_{coupling}
\end{equation}
where the atomic term, in the rotating framce, is given by ($\hbar=1$):

\begin{eqnarray}
H_{atom}&=&\omega_{e}|e\rangle\langle e|+\omega_{g}|g\rangle\langle g|+(\omega_{c}+\omega_s)|s\rangle\langle s|\nonumber\\
&+&\Omega^{*}|s\rangle\langle e|+\Omega|e\rangle\langle s|
\end{eqnarray}
where $\omega_{i}$ is the energy of the state $|i\rangle$ for $i=g,e,s$, and $\omega_c$ is the frequency of the control field.
We assume the waveguide has linear dispersion, so the
Hamiltonian of free propagating photons can be written as:

\begin{equation}
H_{field}=-ic\int_{-\infty}^{+\infty}dx\left(\mathcal{E}_{R}^{\dagger}(x)\frac{\partial}{\partial x}\mathcal{E}_{R}(x)-\mathcal{E}_{L}^{\dagger}(x)\frac{\partial}{\partial x}\mathcal{E}_{L}(x)\right),
\end{equation}
where the annihilation operator for the left- (right-) going field at position $x$ is denoted as
$\mathcal{E}_{L}(x)(\mathcal{E}_{R}(x))$, respectively and $c$ is the group velocity in the waveguide.
Finally, the atom-field coupling is described by:

\begin{equation}
H_{coupling}=-g\sqrt{2\pi}\int_{-\infty}^{+\infty}dx\delta(x-x_{0})|g\rangle\langle e|(\mathcal{E}_{R}^{\dagger}(x)+\mathcal{E}_{L}^{\dagger}(x))+h.c.,
\end{equation}
where the atom is located at $x_{0}$ and the atom-field coupling coefficient
is $g$. Assuming that there is only one excitation inside the system,
the most general state of the system can be written as:

\begin{eqnarray}
|\psi_{k}\rangle&=&\int_{-\infty}^{+\infty}dx\left(\phi_{R}(x)\mathcal{E}_{R}^{\dagger}(x)+\phi_{L}(x)\mathcal{E}_{L}^{\dagger}(x)\right)|g,0\rangle\\
&+&P|e,0\rangle+S|s,0\rangle,
\end{eqnarray}
where the first term in the ket refers to the atomic state and the
second term refers to the photon Fock state. Now, we consider a situation
where the field is propagating to the right and scatter from the atom
at location $x_{0}$. Therefore, the photonic wave function will take
the form
\begin{eqnarray*}
\phi_{R}(x) & = & \theta(-x+x_{0})e^{ikx}+t_a\theta(x-x_{0})e^{ikx},\\
\phi_{L}(x) & = & r_a\theta(-x+x_{0})e^{-ikx},
\end{eqnarray*}
where $t _a(r_a)$ is the atomic transmission (reflection) coefficient, respectively.
By solving the equation $H|\psi_{k}\rangle=\epsilon_{k}|\psi_{k}\rangle$,
we find that:

\begin{eqnarray*}
\epsilon_{k} & = & ck\\
\epsilon_{k}P & = & (\omega_{e}-i\Gamma_{out}/2)P+\Omega S-g\sqrt{2\pi}(1+r_a)\\
-ic(t_a-1) & = & g\sqrt{2\pi}P\\
1+r_a & = & t_a\\
\epsilon_k S &=&(\omega_c+\omega_s) S+\Omega^*P.
\end{eqnarray*}
Note that we have added the effect of the spontaneous emission into the
outside photonic modes by introducing a polarization decay term with
the rate $\Gamma_{out}/2$. We assume the ground state decay is negligible.
Therefore, the reflection coefficient of light scattering at each
site is given by: 
\begin{equation}
r_a  =  -\frac{\Gamma_{1D}\delta}{\delta(\Gamma_{1D}+\Gamma_{out}-2i(\delta+\Delta))+2i|\Omega|^{2}}.\\
\end{equation}
where $\Gamma_{1D}=4\pi g^2/c$ is the spontaneous emission rate into the waveguide. The total spontaneous emission rate is $\Gamma_{tot}=\Gamma_{1D}+\Gamma_{out}$. Therefore, the scattering matrix of a single cell will be:
\begin{equation}
M_{\rm{atom}}=\left(\begin{array}{cc}
(1+i\zeta)e^{ik_{p}a} & i\zeta e^{-ik_{p}a} \\
-i\zeta e^{ik_{p}a} & 1-i\zeta e^{-ik_{p}a} \end{array}\right)\end{equation} 
where $\zeta  =  -\frac{\Gamma_{1D}\delta}{\delta(i\Gamma_{out}+2(\delta+\Delta))-2|\Omega|^{2}}$. By defining the lattice frequency as $\omega_{lattice}=c\frac{2\pi}{a}$, we can write the probe wave number in terms of the detuning from the
atomic transition: $k_{p}a=(k_{p}-k_{a})a+k_{a}a=(\delta+\Delta)\frac{a}{c}+2\pi\frac{\omega_{a}}{\omega_{lattice}}$.
For simplicity, we assume the resonant EIT case
$(\Delta=0)$.

The band structure of a 1D array of 50,000 sites is shown in Fig.\ref{fig:band_theory}.
Due to the atomic periodic dispersion, one can observe the appearance of a band gap and also the finite
size oscillations at the band gap edge \cite{Deutsch:1995,Witthaut:2010}. Note that number used in Fig.\ref{fig:band_theory} are far from accessible regime in the current experiments \cite{Vetsch:2009,Hoffman:2011}.

\begin{figure}
\includegraphics[width=0.4\textwidth]{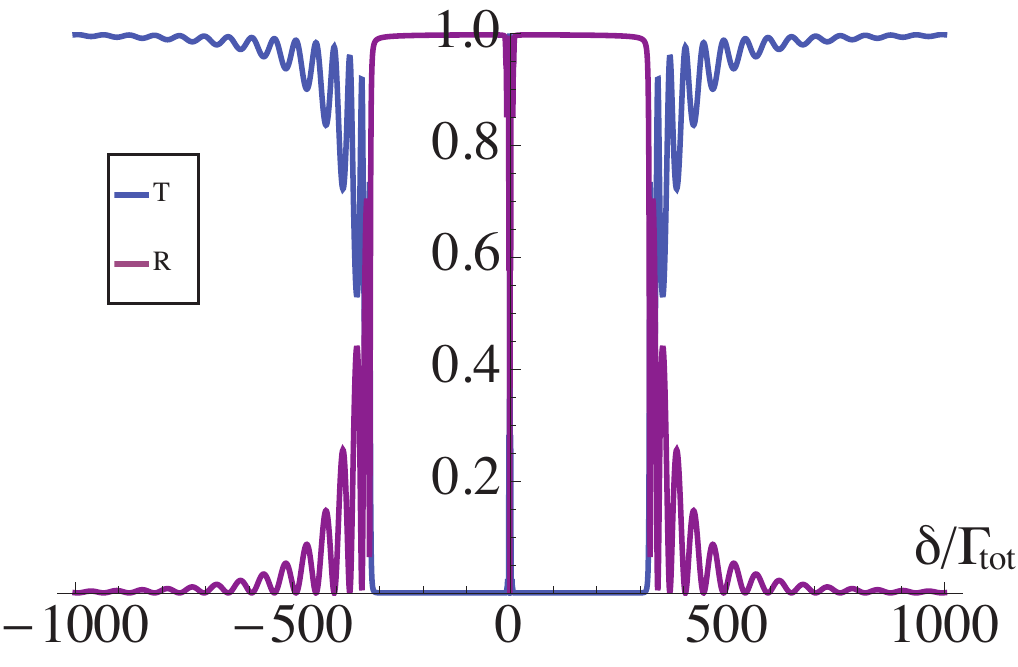}

\caption{Transmission and reflection spectrum. Due to finite size effect, sharp oscillations occur at the band gaps edges. For this plot, a lattice of 50,000 sites is considered, also $\omega_{lattice}=\omega_a$ and $(\Delta,\Gamma_{1D},c/a,\Omega)/\Gamma_{tot}=(0,0.1,10^6,10)$.  \label{fig:band_theory}}
\end{figure}

As mentioned in the main text, for the numbers close to the experimental parameters, we recover a transmission spectrum
similar to that of free space, as shown in Fig.~3 of the main text. There, the dips in the transmission spectrum corresponds to dressed states
split by the Rabi frequency of the control field, in direct analogy
to EIT in free space.

\section*{S2: Adiabatic passage between two LC resonators using the dark state}

\begin{figure}[h]
\center
\includegraphics[width=0.4\textwidth]{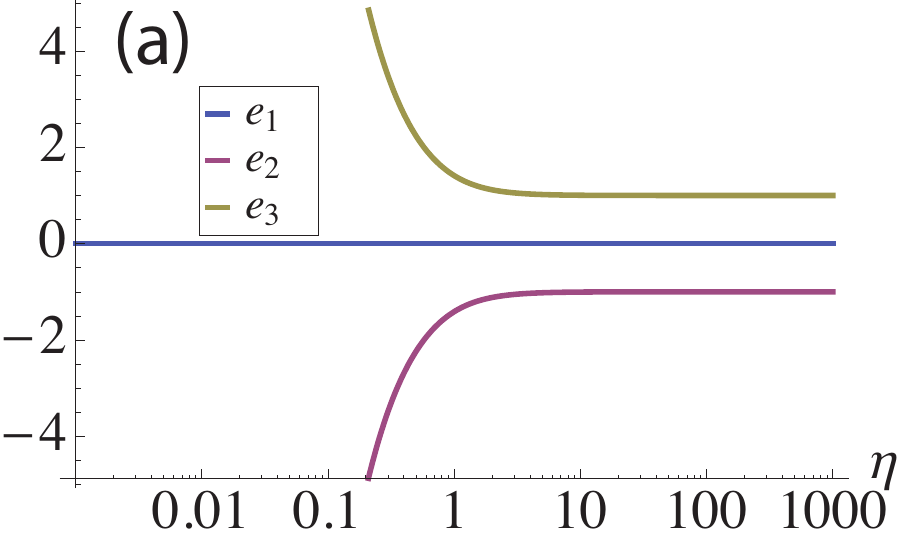}
\includegraphics[width=0.4\textwidth]{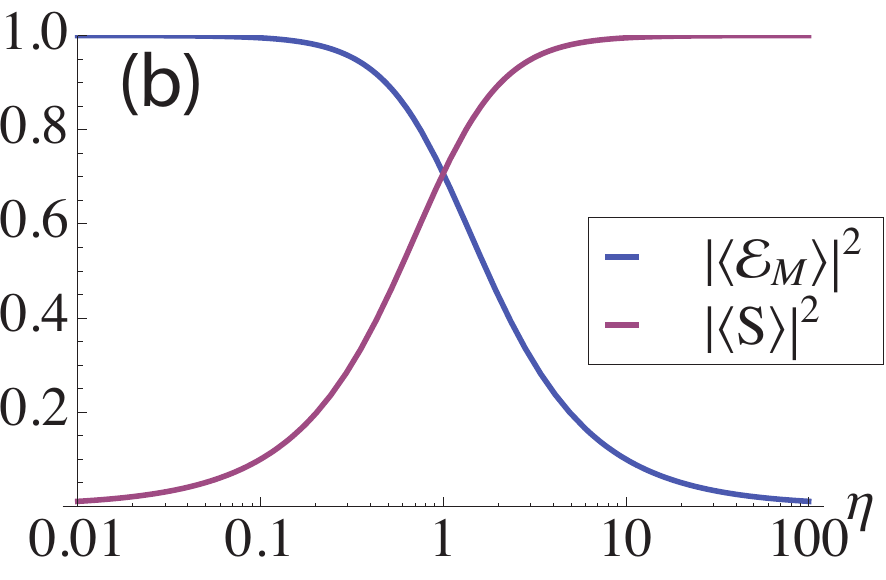}\caption{single cell: (a) energy of the eigenstates in units of $g\sqrt{N}$. Note that the dark state has zero energy and is separated from the other states. (b) The dark state nature changes from photonic to atomic by changing the control field, characterized by $\eta$. \label{fig:single-EIT}}
\end{figure}

In this section, we discuss the adiabatic transfer of a single excitation from one LC resonator to another, using a dark state.
First we consider a single cell where an atomic ensemble is coupled
to a resonator. Since the system is linear, we solve the problem for
a single excitation. The result can be easily generalized to any number
of excitations, as long as the number of excitations remains smaller
than than the number of atoms to guarantee the bosonic commutation
relations \cite{Fleischhauer:2002}. The equations of motion are given
by \cite{Fleischhauer:2000kx,Zimmer:2008}:

\[
i\frac{dX}{dt}=H_{0}X,
\]
where $X=(\langle\mathcal{E}_M\rangle,\langle S \rangle, \langle P \rangle)^{T}$ represents the state of the system: $\langle\mathcal{E}\rangle$ is the mircowave electric field,  $\langle S \rangle $ is the atomic coherence for $|c\rangle \leftrightarrow |a\rangle$ transition, and $\langle P \rangle $ is the atomic coherence for $|b\rangle \leftrightarrow |a\rangle$ transition. The Hamiltonian is 

\begin{equation}
H_{0}=-g\sqrt{N}\left(\begin{array}{ccc}
0 & 0 & 1\\
0 & 0 & \eta^{-1}\\
1 & \eta^{-1} & \delta_2
\end{array}\right)
\end{equation}
where $\eta=\frac{g\sqrt{N}}{\Omega}$ and the dimesionless detuning
is $\delta_2=\frac{\Delta_2}{g\sqrt{N}}.$ We use $g\sqrt{N}$ for the unit of energy. In this case, the group velocity reduction factor
is $\frac{c}{v_{g}}=\sqrt{1+\eta^{2}}.$ When the atmoic medium is
inside a resonator, the slow light effect  manifests itself as narrowing of
the resonator bandwidth \cite{Lukin:98}. The energy spectrum of the system as a function
of the control field -- which is characterized by $\eta$ -- is shown in Fig.~\ref{fig:single-EIT}.
The zero energy state is the dark state $|D\rangle=\frac{1}{\sqrt{1+\eta^{2}}}(|E\rangle-\eta|S\rangle)$, where $|E\rangle=(1,0,0)^T$ represents the photonic state and $|S\rangle=(0,1,0)^T$ represents the atomic coherence. Therefore, one can change the nature of the dark state from photonic to atomic by increasing $\eta$, or equivalently decreasing the magnitude of the control field $\Omega$.

Now, we consider two cells to investigate the adiabatic passage
of a single excitation between them, by simply changing their respective
control field ($\eta_1,\eta_2$). The state of the system can be represented by $X=(X_{1},X_{2})^{T}$,
and the coupling matrix is:

\begin{equation}
H=\left(\begin{array}{cc}
H_{1} & H_{c}\\
H_{c} & H_{2}
\end{array}\right)
\end{equation}
where $H_{i}=H_{0}(\eta\rightarrow\eta_{i})$ and $H_{c}=\left(\begin{array}{ccc}
-\kappa & 0 & 0\\
0 & 0 & 0\\
0 & 0 & 0
\end{array}\right)$ respresents the electric field coupling between resonators. Before
presenting the scheme, we study the behavior of the system for different control field values
$(i.e., \eta_{1},\eta_{2})$. In particular, we investigate the situation
where they are inversely changed: $\eta_{2}=\eta_{c}^{2}\eta_{1}^{-1}$ and the control field values cross at $\eta_{c}$.

\begin{figure}
\includegraphics[width=0.4\textwidth]{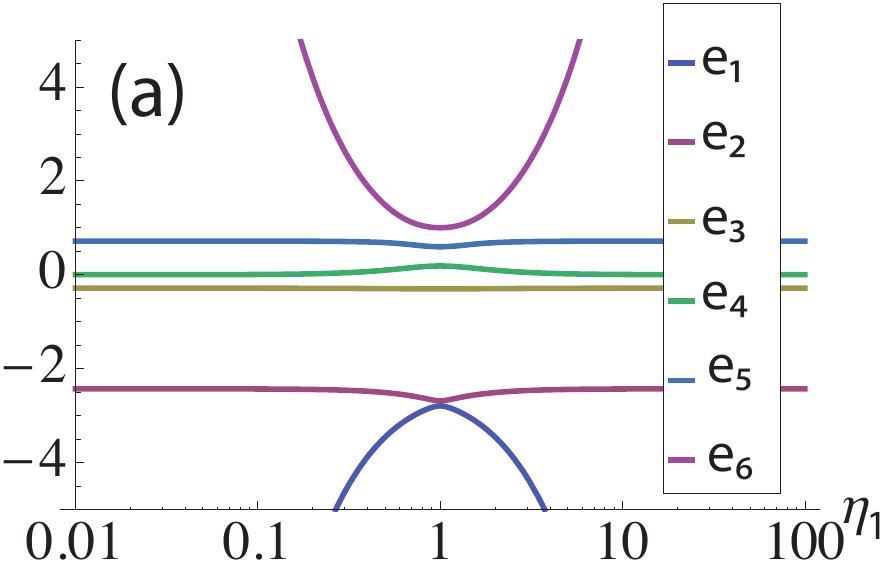}
\includegraphics[width=0.4\textwidth]{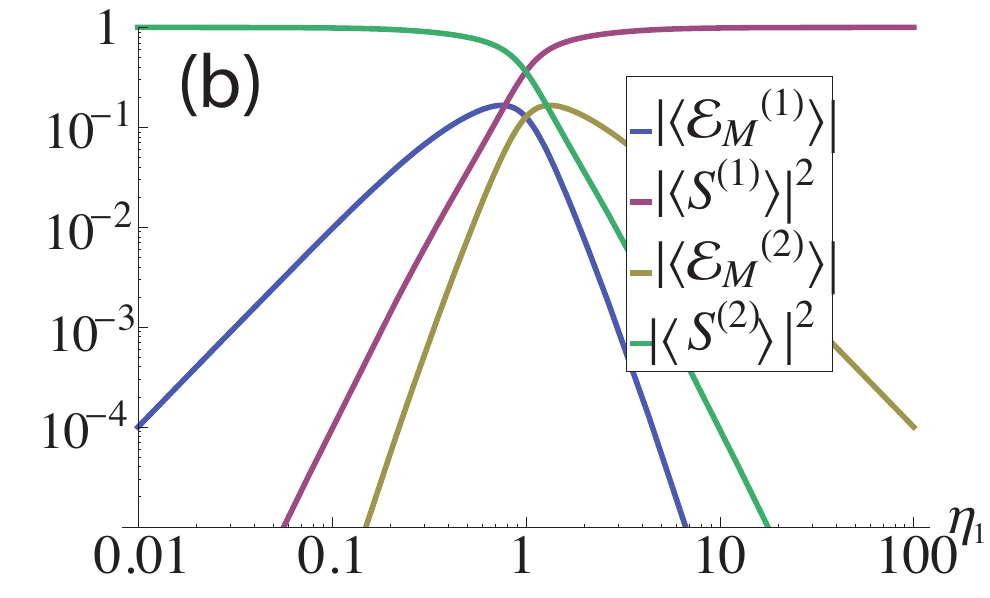}\caption{Double cell: (a) Energy of the eigenstates in units of $g\sqrt{N}$, (b) Probabilities of photonic and atomic components of the forth lowest eigenstate $|e_4\rangle$. By changing the control fields (increasing $\eta_1$ and decreasing $\eta_2$), the atomic coherence transfers from the second cell to the first cell, characterized by $|\langle S^{(2)}\rangle|^2$ and $|\langle S^{(1)}\rangle|^2$, respectively. In these plots, $(\kappa,\delta_2)/g\sqrt{N}=(0.5,2)$ and $\eta_c=1$. \label{fig:double-EIT}}
\end{figure}

The energy eigenstates of the system are detonated by $|e_i\rangle$ with  their corresponding energies $e_i$.  The energy
spectrum of such coupled system is shown in Fig.\ref{fig:double-EIT}.
Due to coupling between the resonators, the dark states are split.
Since the coupling between dark states is given by $\frac{-\kappa}{\sqrt{(1+\eta_{1}^{2})(1+\eta_{2}^{2})}}$,
this splitting is more pronounced when $ $$\eta_{1}\simeq\eta_{2}=1$.
If this was the only coupling mechanics, then at very large or small $\eta$'s,
the splitting would vanish. However, there
is a finite splitting between the new dark states which is constant
in those limits. The latter splitting is due to coupling of the dark
states to the bright and excited states. Using
second order perturbation theory, we find that the energy corrections
to the dark states are equal to $e_{3}\simeq-\delta\kappa^{2},e_{4}\simeq0$
for both large and small $\eta$'s. When two cells are decoupled from each other, if $\eta_{1}\ll1$ (and
$\eta_{2}\gg1$), then we have $|D^{(1)}\rangle\simeq|E^{(1)}\rangle$ and $|D^{(2)}\rangle\simeq|S^{(2)}\rangle$.
Therefore, in this limit, the coupled dark states are: $|e_3\rangle\simeq|E^{(1)}\rangle$
and $|e_4\rangle\simeq|S^{(2)}\rangle$. Similarly, for the opposite limit,
when $\eta_{2}\ll1$ (and $\eta_{1}\gg1$), we will have $|e_3\rangle\simeq|E^{(2)}\rangle$
and $|e_4\rangle\simeq|S^{(1)}\rangle$. 

Therefore, we can adiabatically transfer an atomic excitation from the
resonator 2 to the resonator 1 while remaining in the dark state ($|e_4\rangle$).
The adiabaticity condition is 
 $\sum_{i\neq4} \frac{|\langle
  e_4|\partial_\eta|e_i\rangle|}{e_4-e_i}\dot\eta =\epsilon_1 \dot\eta
\ll 1$, where the sum is over all the states except the dark state of interest. Therefore, a slower transfer is more adiabatic. However, a slower transfer leads to higher loss of the excitation which is characterized by: $\sum_{i=1,2}\int (\kappa_{in}
|\langle\mathcal{E}^{(i)}_M\rangle|^2+\gamma |\langle S^{(i)} \rangle|^2)
\frac{1}{\dot \eta}d\eta = \epsilon_2/ \dot \eta$, where the
first (second) term represents the photonic (atomic) loss,
respectively. The photonic loss is due to the intrinsic loss of the resonator. In general, there should also be a term for the polarization ($|\langle P^{(i)} \rangle|^2$)  loss; however, for the dark state transfer, the probability of the excitation being in the excited state is negligible and therefore one can ignore that the polarization loss.

Combining these two conditions, we find that one should always satisfy $\epsilon_1 \epsilon_2 \ll 1$, regardless of the control field changing rate. We can use this condition to find optimized values of the detuning ($\Delta_2$) and the control field crossing $\eta_c$. An example of this optimization is shown in Fig.~3 of the main text.

\bibliographystyle{apsrev}

\end{document}